\input{aipcheck}

\documentclass[
    ,final            
  ]
  {aipproc}

\layoutstyle{6x9}

\usepackage{xspace}

\def\slash#1{\setbox0=\hbox{$#1$}  
   \dimen0=\wd0     
   \setbox1=\hbox{/} \dimen1=\wd1  
   \ifdim\dimen0>\dimen1   
      \rlap{\hbox to \dimen0{\hfil/\hfil}} 
      #1     
   \else     
      \rlap{\hbox to \dimen1{\hfil$#1$\hfil}} 
      /      
   \fi}      %
\newcommand{\elll}{b}
\newcommand{\tcdot}{{\cdot}}

\newcommand{\TMDPs}{TMD PDFs\xspace}
\newcommand{\TMDFs}{TMD FFs\xspace}

\newcommand{\key}{\mathbf{p}_y}
\newcommand{\Phperp}{\mathbf{P}_{h\perp}}

\newcommand{\mN}{M}
\newcommand{\kei}{{\ensuremath{p}}}
\newcommand{\nn}{\nonumber \\}
\newcommand{\be}{\begin{equation}}
\newcommand{\ee}{\end{equation}}
\newcommand{\bea}{\begin{eqnarray}}
\newcommand{\eea}{\end{eqnarray}}
\newcommand{\zh}{z}
\newcommand{\bm}{\mathbf}
\newcommand{\bpar}{\mathcal{B}_T}

\newcommand{\BMKei}{K}
\newcommand{\BMpei}{p}
\newcommand{\BMpeiT}{\mathbf{\BMpei}_T}

\newcommand{\BMKeiT}{\mathbf{\BMKei}_T}
\newcommand{\Kt}{\BMKeiT}
\newcommand{\BMelll}{b}
\newcommand{\BMelllT}{\mathbf{\BMelll}_T}
\newcommand{\bt}{\BMelllT}
\newcommand{\BMSpin}{S}
\newcommand{\BMSpinT}{\mathbf{\BMSpin}_\perp}

\newcommand{\slim}{\mskip 1.5mu}              
\newcommand{\tAmp}{\ensuremath{\widetilde{A}^{(+)}}}

\newcommand{\Bmp}{\ensuremath{B^{(+)}}}
\begin{document}

\title{Semi-Inclusive Deep Inelastic Scattering and Bessel-Weighted Asymmetries}



\author{Leonard Gamberg}{
  address={Department of Physics, 
Penn State University-Berks, Reading, PA 19610, USA }
}

\author{ Dani\"el Boer}{
  address={Theory Group, KVI, University of Groningen, The Netherlands
}
}

\author{Bernhard Musch}{
  address={Jefferson Lab, Newport News, VA 23606, USA}
}

\author{Alexei Prokudin}{
  address={Jefferson Lab, Newport News, VA 23606, USA}
 }

\begin{abstract}
We consider the cross section in Fourier space, 
conjugate to the outgoing hadron's transverse momentum, 
where convolutions of transverse momentum dependent parton distribution 
functions  and fragmentation functions become simple products.  Individual asymmetric terms in the cross section can be projected out by means of a
generalized set of weights involving Bessel functions. Advantages of employing these 
Bessel weights are that they suppress (divergent) contributions from high transverse 
momentum and that soft factors cancel in (Bessel-) weighted asymmetries.
Also, the resulting 
compact expressions immediately connect to previous work on evolution equations for transverse 
momentum dependent parton distribution and fragmentation functions and to quantities accessible 
in lattice QCD. Bessel-weighted asymmetries are thus model independent observables that augment 
the description and our understanding of correlations of spin and momentum in nucleon structure.

\end{abstract}

\maketitle


\subsection{Introduction}
\begin{figure}[t]
	\centering
	\includegraphics[width=0.4\textwidth]{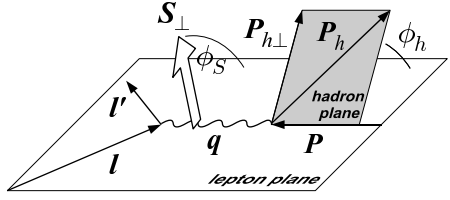}
	\caption{Kinematics of the SIDIS process. The in- and out-going lepton momenta are $l$ and $l'$, respectively. The momentum transfer is $q$.
	The target nucleon carries momentum $P$ and its transverse spin components are labelled $\BMSpinT$. The momentum of the measured hadron $P_h$ has transverse components $\Phperp$, which define an angle $\phi_h$ with the lepton plane. \label{fig-kin}}
	\end{figure}

In the factorized picture of semi-inclusive processes,  
where the transverse momentum of the detected hadron  $\Phperp$ is small compared to the photon virtuality $Q^2$,  
transverse momentum dependent (TMD) parton distribution functions (PDFs) characterize 
the spin and momentum structure of the proton 
\cite{Ralston:1979ys,Sivers:1989cc,Collins:1992kk,
Kotzinian:1994dv,Mulders:1995dh,Kotzinian:1997wt,Boer:1997nt}.
At leading twist there are 8 \TMDPs. 
They can be studied experimentally by analyzing angular modulations 
in the differential cross section, so called 
spin and azimuthal asymmetries. These modulations are a function of the azimuthal angles of 
the final state hadron momentum about
the virtual photon direction,  as well as that of the target polarization (see 
Fig.~\ref{fig-kin} and e.g., Ref.~\cite{Bacchetta:2006tn} for a review). 
 \TMDPs enter the SIDIS cross section in momentum space 
convoluted with transverse momentum dependent fragmentation functions (\TMDFs).  However, after a two-dimensional Fourier transform of the cross section with respect to the transverse hadron momentum $\Phperp$,
these convolutions become simple products of functions in Fourier $\BMelllT$-space.   
The usefulness of Fourier-Bessel transforms in 
studying the factorization as well as the scale dependence of 
transverse momentum dependent cross section
has been known for some time~\cite{Parisi:1979se,Jones:1980my,Collins:1981uw,Ellis:1981sj,
Collins:1984kg,Ji:2004xq,CollinsBook}.   
Here we exhibit the  structure of the cross section in  $\BMelllT$-space
and demonstrate how this representation results in 
model independent observables which are generalizations of 
the conventional weighted 
asymmetries~\cite{Kotzinian:1995cz,Kotzinian:1997wt,Boer:1997nt}.  
Additionally, we explore the impact that these observables have in studying
the scale dependence of the SIDIS cross section at small to  moderate 
transverse momentum where the TMD framework is designed to give a good 
description of the cross section.  In particular we study how the so called
soft factor cancels from these observables. 
The soft factor~\cite{Collins:1999dz,CollinsBook,Collins:2004nx,Ji:2004xq,Ji:2004wu,Aybat:2011zv} is an essential  element of the cross section  that emerges in the proofs of  TMD factorization~\cite{Collins:1981uw,Collins:1984kg,Ji:2004xq,CollinsBook}. It 
accounts for  the collective effect of soft momentum gluons not 
associated with either the distribution or fragmentation part of the  process 
and it is shown to be universal in hard processes \cite{Collins:2004nx}. 

\subsection{Weighted asymmetries}

The concept of transverse momentum 
weighted Single Spin Asymmetries (SSA) was proposed some time ago in  
Ref.~\cite{Kotzinian:1997wt,Boer:1997nt}.  
Using the technique of weighting enables one 
to disentangle in a model independent way the cross sections 
and asymmetries in terms of the transverse momentum moments of TMDs. A comprehensive  
list of such weights was derived in Ref.~\cite{Boer:1997nt} for 
 semi-inclusive deep inelastic scattering (SIDIS).
In SIDIS  and Drell-Yan scattering, proofs 
of TMD factorization contain an additional factor, the soft factory~\cite{Collins:1999dz,CollinsBook,Collins:2004nx,Ji:2004xq,Ji:2004wu,Aybat:2011zv}.  
At tree level (zeroth order in $\alpha_S$) the soft factor is unity, which explains its absence in the factorization formalism 
considered for example in Ref.~\cite{Bacchetta:2006tn}. Consequently it is also absent   in tree level phenomenological analyses of the experimental 
data (see for example  
Refs.~\cite{Anselmino:2005nn,Collins:2005ie,Anselmino:2007fs}).  
Potentially the results of  analyses at  different energies
can be difficult to compare . For a correct description of the energy scale dependence of the cross sections
and asymmetries involving TMDs, the soft factor is essential to include~\cite{Ji:2004wu,Ji:2004xq,Idilbi:2004vb}. However, it is possible to consider observables where the soft factor is indeed 
absent or cancels out. These are precisely the weighted asymmetries.  

We will focus on the Sivers asymmetry~\cite{Sivers:1989cc}.  With a general  $|\mathbf{P}_{h\perp}|$-weight, this asymmetry can be written as~\cite{Boer:1997nt}
\bea
	A^{w_1 \sin(\phi_h-\phi_S)}_{UT}(x,z,y) &\equiv &  \nn
&&\hspace{-3.5cm}\frac{2\int d |\mathbf{P}_{h\perp}|\, d \phi_h\, d \phi_S\, w_1(|\mathbf{P}_{h\perp}|)\ \sin(\phi_h - \phi_S)\ ( d\sigma(\phi_h,\phi_S) - d\sigma(\phi_h,\phi_S+\pi )) }
	            {\int d |\mathbf{P}_{h\perp}|\, d \phi_h\, d \phi_S\,  w_0(|\mathbf{P}_{h\perp}|)\  ( d\sigma(\phi_h,\phi_S) + d\sigma(\phi_h,\phi_S+\pi ) )}.
	\label{eq:BMwtasym}      
\eea
The above asymmetry has already been discussed for the choice 
$w_1 =\frac{|\mathbf{P}_{h\perp}|}{Mz}$, $w_0=1$.  
In the numerator, the angular weight $\sin(\phi_h-\phi_S)$ projects out the structure function $F_{UT,T}^{\sin(\phi_h-\phi_S)}$~\cite{Bacchetta:2006tn}
from the cross section, while the $ |\mathbf{P}_{h\perp}|$-weight leads to a ``deconvolution'' of that structure function into a product of the first $\BMpeiT$-moment of the Sivers function $f_{1T}^{\perp(1)}(x)$ and the lowest moment of the unpolarized fragmentation function $D_1^{(0)}(z)$~\cite{Boer:1997nt}.

In~\cite{Boer:2011xd}, we  show  that in TMD factorization
 the soft factor cancels in the asymmetry, due to the ``deconvolution'' achieved by appropriate $ |\mathbf{P}_{h\perp}|$-weighing. We demonstrate that it is a natural generalization to employ Bessel functions as weights, $w_{n} \propto J_n(|\mathbf{P}_{h\perp}| \mathcal{B}_T)$, where $\mathcal{B}_T$ is the Fourier conjugate variable
of $|\mathbf{P}_{h\perp}|$.

Moreover, this generalization addresses a problem related to the perturbative tail of TMDs.  
Without further regularization, the integrals defining the moments $f_{1T}^{\perp(1)}(x)$ and $f^{(0)}(x)$ are ill-defined due to the asymptotic behavior 
$f_{1T}^{\perp}(x,\BMpeiT^2) \propto 1/\BMpeiT^4$,  $f_{1}(x,\BMpeiT^2) \propto 1/\BMpeiT^2$ at large $\BMpeiT$ (see~\cite{Bacchetta:2008xw}). Using Bessel functions as weights, the respective integrals become convergent while the ``deconvolution'' property and soft factor cancellation are preserved. However, it is important to stress that while the integrals are
convergent, the scale dependence of the resulting functions, and consequently also the $Q^2$ dependence of the asymmetries, 
remains to be studied. Also the link with gluonic pole matrix elements~\cite{Boer:2003cm} becomes less direct.

\subsection{Bessel-weighted asymmetries} 

To deconvolute or convert the convolutions of \TMDPs and \TMDFs in the SIDIS cross section into products, one can perform a multipole expansion and a subsequent Fourier transform of the cross section with respect to the transverse components $\Phperp$ of the hadron momentum. 
In general, we can write a transverse momentum dependent cross section $\sigma(|\Phperp|,\phi_h)$ in the form
\bea
  \sigma(|\Phperp|,\phi_h) & = &
  \int \frac{d^2 \BMelllT}{(2\pi)^2}\  e^{-i \Phperp\cdot \BMelllT}\ \tilde \sigma(\BMelllT) \nonumber \\
  & = & \int \frac{d |\BMelllT|}{2\pi}\, |\BMelllT| \int_0^{2\pi} \frac{d \phi_\elll}{2\pi}\  e^{-i |\Phperp| |\BMelllT| \cos(\phi_h - \phi_\elll) }\ \sum_{n=-\infty}^\infty e^{i n \phi_\elll}  \tilde \sigma_n(|\BMelllT|) \nonumber \\
  & = &\sum_{n=-\infty}^\infty  e^{i n \phi_h} \int \frac{d |\BMelllT|}{2\pi} \, |\BMelllT|\ 
  (-i)^n  J_n(|\Phperp| |\BMelllT|)\  \tilde \sigma_n(|\BMelllT|) \, ,
  \eea
where $\tilde \sigma(\BMelllT) =  \sum_{n=-\infty}^\infty e^{i n \phi_\elll}  \tilde \sigma_n(|\BMelllT|)$ is a two-dimensional multipole expansion of the cross section in Fourier space.
The $n^{\rm th}$ harmonic in $\phi_h$ is accompanied by the $n^{\rm th}$ Bessel function of the first kind $J_n$.

With these definitions, the relevant terms of the SIDIS cross section 
can be written as
\bea
\frac{d\sigma}{dx \, dy\, d\phi_S \,dz\, d\phi_h\, |\mathbf{P}_{h\perp}|\, d |\mathbf{P}_{h\perp}|} &=& \frac{\alpha^2}{x y\slim Q^2}
\frac{y^2}{(1-\varepsilon)}  \biggl( 1+\frac{\gamma^2}{2x} \biggr) 
  \int_0^\infty \frac{d |\BMelllT|}{(2\pi)} |\BMelllT| \nn
&&\hspace{-5cm}\times \Biggl\{
J_0(|\BMelllT|\,|\mathbf{P}_{h\perp}| )\, \widetilde{F}_{UU ,T} 
\ +\ 
|\BMSpinT| \sin(\phi_h-\phi_S)\, J_1(|\BMelllT|\,|\mathbf{P}_{h\perp}| )\,  \widetilde{F}_{UT ,T}^{\sin(\phi_h -\phi_S)}  \, +\,  \ldots 
\Biggl\}, \nn
	\label{eq:BMcrosssec}
\eea
where the ellipsis represent 16 more terms.  Note that there is only a finite number of multipoles in the SIDIS cross section; the Bessel function of highest order is $J_3$. Here we show only two terms in the cross section and omit for now regularization parameters needed beyond tree level, see Ref. \cite{Boer:2011xd} and references therein for more details.  Introducing the  Fourier-transformed TMDs and fragmentation functions
\bea
	\tilde f(x, \BMelllT^2 ) & \equiv& \int d^2 \BMpeiT\ e^{i \BMpeiT \cdot \BMelllT} \ f(x, \BMpeiT^2 ) \nonumber  =  2\pi \int_0^\infty d|\BMpeiT|\, |\BMpeiT|\, J_0(|\BMelllT|\,|\BMpeiT|) \, f(x, \BMpeiT^2 )\ , \label{eq-bttransf} \\
\tilde D(z, \bt^2) & \equiv& \int d^2 \Kt \,  e^{i  \bt\cdot \Kt }\; D(z, \Kt^2) 
	= 2\pi\int  d|\Kt| |\Kt|\ J_0(|\bt||\Kt|)\  D(z, {K_T^2} )\; .
%
\eea
and the derivative (or $\BMelllT$ moment)
\bea
	\tilde f^{(1)}(x, \BMelllT^2 ) \equiv \ - \frac{2}{M^2}\,  \partial_{\BMelllT^2}\, \tilde f(x, \BMelllT^2 ) = \frac{2\pi}{M^2} \int_0^\infty d|\BMpeiT|\, \frac{|\BMpeiT^2|}{|\BMelllT|}\, J_1(|\BMelllT|\,|\BMpeiT|)\  f(x, \BMpeiT^2 ) \, ,
\label{eq-btder}
\eea
the structure functions in Fourier space in 
Eq.~(\ref{eq:BMcrosssec}) are given by
\bea
	\widetilde{F}_{UU ,T} & = & x\sum_a e_a^2\ \tilde f_{1}^{(0)a}(x,z^2\BMelllT^2) \ \tilde D_{1}^{(0)a}(z,\BMelllT^2)\ \tilde S(\BMelllT^2)\ H_{UU,T}(Q^2)\, , \label{eq:BMtildeFUUT} \\
	\widetilde{F}_{UT ,T}^{\sin(\phi_h -\phi_S)} & = & -x\sum_a e_a^2 |\BMelllT| z M\,\tilde f_{1T}^{\perp(1)a}(x,z^2\BMelllT^2)  \tilde D_{1}^{(0)a}(z,\BMelllT^2)\ \tilde S(\BMelllT^2)\, H_{UT,T}(Q^2)\, ,
	\label{eq:BMtildeFUTT}
\eea
where, $\tilde f_{1}^{(0)a}$ and $\tilde f_{1T}^{\perp(1)a}$ are the 
Fourier transformed unpolarized and Sivers \TMDPs respectively, 
and  $\tilde D_{1}^{(0)a}$ is a Fourier transformed fragmentation function.  
We have used the kinematic variables 
$Q^2 \equiv -q^2$, $M^2=P^2$, $x\approx x_B \equiv Q^2/P\cdot q$, $y = P\cdot q/P \cdot l$, and $z \approx z_h \equiv P\cdot P_h / P \cdot q$
and assume $M \ll Q$, $|\mathbf{P}_{h\perp}| \ll z Q$. The sum $\sum_a$ runs over quark flavors $a$ and $e_a$ is the corresponding electric charge of the quark.  In contrast to the tree-level equation~\cite{Bacchetta:2006tn,Boer:2011xd}, 
we include here an explicit soft factor $\tilde S(\BMelllT^2)$ (in Fourier space) and a hard part $H(Q^2)$, as in  reference~\cite{Ji:2004wu,Ji:2004xq}. For brevity we suppress the
 dependencies on a UV cutoff scale $\mu$ and on rapidity cutoff parameters (e.g., $\zeta$, $\hat \zeta$, $\rho$ in~\cite{Ji:2004wu,Ji:2004xq,Boer:2011xd}).
It is now clear that the cross section equation~\eqref{eq:BMcrosssec} is a multipole series, and that projection on Fourier modes in polar coordinates $\phi_h$, $|\mathbf{P}_{h\perp}|$ will give access to the right hand sides of equations~\eqref{eq:BMtildeFUUT} and \eqref{eq:BMtildeFUTT}. Calculating the weighted asymmetry equation~\eqref{eq:BMwtasym} with weights $w_1(|\mathbf{P}_{h\perp}|) = 2 J_1( |\mathbf{P}_{h\perp}| \mathcal{B}_T ) / z M \mathcal{B}_T $ and 
$w_0(|\mathbf{P}_{h\perp}|) = J_0( |\mathbf{P}_{h\perp}| \mathcal{B}_T )$ thus 
yields
\bea
 A_{UT}^{\frac{{2 J}_1(|\bm{P}_{h\perp}|\bpar )}{\zh M \bpar}\sin(\phi_h - \phi_s)}(x,y,z;\bpar) 
& =\  -2
\frac{\sum_a e_a^2\,H_{UT,T}(Q^2)\,  \tilde f_{1T}^{\perp(1)a}(x,z^2 \bpar^2)\,  \tilde S(\bpar^2)\, \tilde D_{1}^{(0)a}(z,\bpar^2)}{
\sum_a e_a^2\, H_{UU,T}(Q^2)\, \tilde f_{1}^{(0)a}(x,z^2 \bpar^2)\, \tilde S(\bpar^2)\, \tilde D_{1}^{(0)a}(z,\bpar^2) 
}\, .
\label{eq:ssa_sivers_final}
\eea
Due to the ``deconvolution'' of the structure functions 
in the weighted asymmetries,  and
universality of the soft factor,  $\tilde{S}$ 
cancels in the numerator and the denominator. 

Further, note that $\bpar$ enters the weights $w_0$ and $w_1$ as a free parameter that we can scan over a whole range in order to compare the transverse momentum dependence of the distributions in the numerator and denominator relative to each other (in Fourier space). At the operator level, $\bpar(=|\BMelllT|)$
 controls the space-like distance between quark fields in the correlation functions we measure.
The Bessel-weighted asymmetries are a natural extension of conventional weighted asymmetries \cite{Kotzinian:1997wt,Boer:1997nt} with weights $w_1$ proportional to powers of $|\Phperp|$. Indeed, in the limit $ \bpar \rightarrow 0$, 
equation~(\ref{eq:ssa_sivers_final}) results in the  often encountered 
special case  for the SIDIS cross section
\bea
A_{UT}^{\frac{|\bm{P}_{\perp}|}{\zh M}\sin(\phi_h - \phi_s)}(x, \zh, y) = 
 -2 \frac{\sum_a e_a^2\ H_{UT,T}(Q^2)\  f_{1T}^{\perp (1) a}(x) \ D_1^{(0)a}(z) 
}{
\sum_a e_a^2\ H_{UU,T}^{}(Q^2)\ f_{1}^{(0)a}(x)\ D_{1}^{(0)a}(z)
}\;.
\label{eq:ssa_sivers_finalbpar0}
\eea
In particular,  we obtain formally the relations $\tilde f_1^{(0)}(x,0) =  f_1^{(0)}(x)$, $\tilde f_{1T}^{\perp(1)}(x,0) =  f_{1T}^{\perp(1)}(x)$ and $\tilde D_1^{(0)}(x,0) =  D_1^{(0)}(x)$, where
\bea
	\tilde f^{(n)} (x,0) = f^{(n)}(x) \equiv \int d^2 \BMpeiT \left( \frac{
\BMpeiT^2}{2 M^2} \right)^n f(x,\BMpeiT^2) \, .
	\label{eq-tmmoment}
\eea
An analogous equation holds for the fragmentation functions $D$.
However, we caution the reader that these moments are not well-defined
without some regularization.  It is therefore safer to study Bessel-weighted asymmetries at finite $\bpar$, where the Bessel functions suppress contributions from large transverse momenta.  

\subsection{Cancellation of soft factor at the level of matrix elements}
In a similar manner to the discussion above,  we now consider the soft factor cancellation in the average transverse momentum shift of unpolarized quarks in a transversely polarized nucleon for a given longitudinal momentum fraction $x$. 
This shift is considered in \cite{Burkardt:2003uw} and  defined 
by a ratio of the $\BMpeiT$-weighted correlator,
\bea
	\langle \kei_y(x) \rangle_{TU} &= \left. \frac{ \int d^2 \BMpeiT
\, \mathbf{p}_y  \ \Phi^{(+)[\gamma^+]}(x,\BMpeiT,P,S,\mu^2,\zeta,\rho) }{ \int d^2 \BMpeiT  \phantom{\mathbf{\kei}_y}\ \Phi^{(+)[\gamma^+]}(x,\BMpeiT,P,S,\mu^2,\zeta,\rho) } \right|_{S^\pm=0,\,\mathbf{S}_T=(1,0)} 
	= \mN \frac{ f_{1T}^{\perp(1)}(x;\mu^2,\zeta,\rho) }{ f_1^{(0)}(x;\mu^2,\zeta,\rho)} \, , 
	\label{eq:ktmomratio}
\eea
where $ f_{1T}^{\perp(1)}$ and $f_1^{(0)}$ are the moments defined in Eqs.\ \eqref{eq-tmmoment}. Obviously, the average momentum shift is very similar in structure to the weighted asymmetry Eq.\ \eqref{eq:ssa_sivers_final}. While the weighted asymmetries are accessible directly from the $\Phperp$-weighted cross section, the average transverse momentum shifts are obtained from the $\BMpeiT$-weighted correlator and could in principle be accessible from weighted jet asymmetries. As already mentioned, the integrals defining the moments of \TMDPs on the right hand side of the above equation are divergent without suitable regularization. Therefore we generalize the above quantity, weighting with Bessel functions of $|\BMpeiT|$. In particular, we replace
\begin{equation}
	\key = |\BMpeiT| \, \sin(\phi_\kei) \quad \longrightarrow \quad 
  \frac{2 J_1( |\BMpeiT| \bpar )}{\bpar} \, \sin(\phi_\kei-\phi_S ) \, ,
\end{equation}
where $\phi_S=0$  for the choice $\mathbf{S}_T=(1,0)$ in Eq. \eqref{eq:ktmomratio}. The correlator $\Phi^{(+)[\gamma^+]}$ reads in terms of the amplitudes $\tAmp_i$
and $\Bmp_i$~\cite{Boer:2011xd}, 
\bea
	\Phi^{(+)[\gamma^+]}(x,\BMpeiT,P,S,\mu^2,\zeta,\rho) & = &
\int \frac{d(\elll \tcdot P)}{(2\pi)} \ e^{ix(\elll \tcdot P)}\,
\int_0^\infty \frac{d |\BMelllT|}{2\pi} |\BMelllT| \ 
\Big\{\  \nonumber \\
	&&\hspace{-5cm}	J_0(|\BMelllT|\, |\BMpeiT|)\, 2\tAmp_{2B}/\widetilde{S} 
- M  |\BMelllT|\, |\mathbf{S}_T|\, \sin(\phi_\kei-\phi_S)\, J_1(|\BMelllT|\, |\BMpeiT|)\,  2\tAmp_{12B}/\widetilde{S} 
	\ \Big\}\, .
\nn
\eea
The Bessel-weighted analog of Eq.\ \eqref{eq:ktmomratio} is thus
\bea
	\langle \kei_y(x) \rangle_{TU}^{\bpar}
	& \equiv \ \left. \frac{
	  \int d|\BMpeiT|\, |\BMpeiT| \int d \phi_\kei
\frac{2\, J_1( | \BMpeiT| \bpar )}{\bpar}
\sin(\phi_\kei - \phi_S)\ \Phi^{(+)[\gamma^+]}(x,\BMpeiT,P,S,\mu^2,\zeta,\rho) }
	{ \int d|\BMpeiT|\, |\BMpeiT| \int d \phi_\kei
J_0( | \BMpeiT| \bpar )\, \phantom{\sin(\phi_\kei - \phi_S) } \Phi^{(+)[\gamma^+]}(x,\BMpeiT,P,S,\mu^2,\zeta,\rho) } \right|_{|\mathbf{S}_T|=1} \nonumber \\
	& \hspace{-2cm}= \mN \frac{ \tilde f_{1T}^{\perp(1)}(x,\bpar^2;\mu^2,\zeta,\rho) }{ \tilde f_1^{(0)}(x,\bpar^2;\mu^2,\zeta,\rho)} \, .
	\label{eq:ktmomratio_Bessel}	
\eea
Again, the soft factors cancel where the independence of the soft factor on $v \tcdot \elll / \sqrt{v^2}$ is crucial~\cite{Boer:2011xd}. Further, 
weighting with Bessel functions at various lengths $\bpar$ thus allows us to map out, e.g., ratios of Fourier-transformed TMDs~\cite{Boer:2011xd}.  In the limit $\bpar \rightarrow 0$, we recover equation \eqref{eq:ktmomratio}, $\langle \kei_y(x) \rangle_{TU}^0 =  \langle \kei_y(x) \rangle_{TU}$ , which we have thus shown to be formally free of any soft factor contribution. However, we caution the reader again that the expressions at $\bpar = 0$ can be ill-defined without an additional regularization step. 

\subsection{Conclusions}

We have demonstrated that rewriting the SIDIS cross-section in coordinate space 
displays the important feature that structure functions become simple products of  
Fourier transformed \TMDPs\ and FFs, or derivatives thereof.  The angular structure of the cross section naturally suggests weighting with Bessel functions in order to project out these 
Fourier-Bessel transformed distributions, which serve as well-defined replacements of the transverse moments
entering conventional weighted asymmetries. In addition, Bessel-weighted asymmetries 
provide a unique opportunity to study nucleon structure in a model independent way due 
to the absence of the soft factor  which as we have shown cancels from these observables.  This cancellation is based on the fact that the soft factor 
is flavor blind in hard processes, and it depends only on $\bm{\elll}^2_T, \mu^2, \rho$. Moreover, evolution equations for the distributions are typically calculated in terms of the (derivatives of) Fourier transformed \TMDPs and FFs. As a result the study of the scale dependence of Bessel-weighted asymmetries should prove more straightforward.  For the above stated reasons we propose Bessel-weighted asymmetries as clean observables to study the scale 
dependence of \TMDPs and FFs at existing (HERMES, COMPASS, JLab) and future facilities (Electron Ion Collider, JLab 12 GeV). 



\begin{theacknowledgments}
LG, thanks the organizers of PacSpin
2011 held in Cairns, Australia, 
 in particular Tony Thomas, for the invitation to present this work. This work was supported in part by the US Department of Energy under contract DE-FG02-07ER41460 (L.G.), 
and is authored by Jefferson Science Associates, LLC under U.S. DOE Contract No. DE-AC05-06OR23177. 
\end{theacknowledgments}



\bibliographystyle{aipproc}   

\bibliography{biblio-2}


\end{document}